\documentclass{PoS}
\usepackage{graphicx}
\usepackage{epstopdf}
\def\be{\begin{equation}}
\def\ee{\end{equation}}
\def\bea{\begin{eqnarray}}
\def\eea{\end{eqnarray}}
\def\ba{\begin{array}}
\def\ea{\end{array}}
\def\bc{\begin{center}}
\def\ec{\end{center}}
\title{Impact parameter space and transverse distortion}

\ShortTitle{Impact parameter space and transverse distortion}

\author{\speaker{Harleen DAHIYA}\thanks{Authors would like to thank S.J. Brodsky for helpful discussions and Department of Science and Technology, Government of India for financial support.}\\
        Department of Physics, Dr. B.R. Ambedkar National
Institute of Technology, Jalandhar, 144011, India \\
        E-mail: \email{dahiyah@nitj.ac.in}}

\author{Narinder KUMAR\\
       Department of Physics, Dr. B.R. Ambedkar National
Institute of Technology, Jalandhar, 144011, India}
\abstract{We investigate the GPDs in impact parameter space using the explicit light front wave functions (LFWFs) for the two-particle Fock state of the electron in QED. The Fourier transform (FT) of the GPDs gives the distribution of quarks in the transverse plane for zero longitudinal momentum transfer ($\zeta=0$). We study the relationship of  the spin flip GPD with the distortion of unpolarized quark distribution in the transverse plane when the target nucleon is transversely polarized and also determine the sign of distortion from the sign of anomalous magnetic moment.}

\FullConference{XXII. International Workshop on Deep-Inelastic Scattering and Related Subjects,\\
		28 April - 2 May 2014\\
		Warsaw, Poland}

\begin{document}
\section{Introduction}
Deep virtual compton scattering (DVCS) \cite{dvcs,dvcs1,dvcs2,dvcs3} is the main process to probe the internal structure of hadrons.  Recently, the Generalized Parton Distributions (GPDs) \cite{gpds,gpds1,gpds2,gpds3,gpds4} have attracted a considerable amount of interest towards this. GPDs  allow us to access partonic configurations not only with a given longitudinal momentum fraction but also at a specific (transverse) location inside the hadron. GPDs can be related to the angular momentum carried by quarks in the nucleon and the distribution of quarks can be described in the longitudinal direction as well as in the impact parameter space \cite{longit,imp2,imp1,imp}. When integrated over $x$ the GPDs reduce to the form factors which are the non-forward matrix element of the current operator and they describe how the forwards matrix element (charge) is distributed in position space. On the other hand, Fourier transform (FT) of GPDs w.r.t. transverse momentum transfer gives the distribution of partons in transverse position space \cite{imp2,imp1}. Therefore, their should be some connection between transverse position of partons and FT of GPDs w.r.t. transverse momentum transfer.

With the help of impact parameter dependent parton distribution function (ipdpdf) one can obtain the transverse position of partons in the transverse plane. However, it is not possible to measure the longitudinal position of partons. In order to measure the transverse position with the  longitudinal momentum simultaneously we can consider the polarized nucleon state in the transverse direction which leads to  distorted unpolarized ipdpdf in the tranverse plane \cite{ipd,ipd1}. Distortion obtained in the transverse plane also leads to single spin asymmetries (SSA) and it has been shown that such asymmetries can be explained by final state interactions (FSI) \cite{ssa,ssa1,fsi}. This mechanism gives us a good interpretation of SSAs which arises from the asymmetry (left-right) of quarks distribution in impact parameter space.

To study the GPDs, we use light front wave functions (LFWFs) which give a very simple representation of GPDs.
Impact parameter dependent parton distribution functions have been investigated by using the explicit LFWFs for the two-particle Fock state of the electron in QED \cite{model,model1}. In our case we take $\zeta=0$ \cite{ipd2,ipd3} which represents the momentum transfer exclusively in transverse direction leading to the study of ipdpdfs in transverse impact parameter space. For the case of spin flip GPD $E(x,0,-\vec{\Delta}_\perp^2)$, the parton distribution is distorted in the transverse plane when the target has a transverse polarization and when integrated over $x$, $E(x,0,-\vec{\Delta}_\perp^2)$ yields the Pauli form factor $F_2(t)$.  The sign of distortion can be concluded from the sign of the magnetic moment of the nucleon. We extend the calculations to unintegrated momentum space distribution which is the another direct way to determine the sign of distortion in impact parameter space from the LFWFs.
\section{Generalized Parton Distributions (GPDs) }
The GPDs $H,E$ are defined through matrix elements of the bilinear vector currents on the light cone:
\bea
&& \int \frac{dy^-}{8 \pi} e^{i x P^+ y^-/2} \langle P'|\bar{\psi}(0) \gamma^+ \psi(y)| P\rangle|_{y^+=0,y_\perp=0} \nonumber\\
&=&\frac{1}{2 \bar{P^+}} \bar{U}(P')[H(x,\zeta,t) \gamma^+ +E(x,\zeta,t) \frac{i}{2M} \sigma^{+ \alpha}(-\Delta_\alpha)] U(P).
\label{e1}
\eea
Here, $\bar{P}=\frac{1}{2}(P'+P)$ is the average momentum of the initial and final hadron, $\zeta$ is the skewness parameter and $t= - \vec{\Delta}_\perp^2$ is the invariant momentum transfer. Since we are considering the case where momentum transfer is purely transverse, we take the skewness parameter $\zeta=0$.

The off-forward matrix elements can be expressed as overlaps of the light front wave functions (LFWFs) for the two-particle Fock state of the electron in QED. We consider here a spin-$\frac{1}{2}$ system as a composite of spin-$\frac{1}{2}$ fermion and spin-1 vector boson. The details of the model have been presented in Ref. \cite{model1}. The helicity non-flip and flip GPDs in this model can be expressed as
\bea
H(x,0,-\vec{\Delta}_\perp^2)&=& \int{\frac{d^2\vec{k}_\perp}{16\pi^3}} \bigg[ \psi_{+\frac{1}{2}+1}^{\uparrow *}(x,\vec{k}'_\perp) \psi_{+\frac{1}{2}+1}^{\uparrow}(x,\vec{k}_\perp)+\psi_{+\frac{1}{2}-1}^{\uparrow *}(x,\vec{k}'_\perp) \psi_{+\frac{1}{2}-1}^{\uparrow}(x,\vec{k}_\perp)+\nonumber\\
&& \psi_{-\frac{1}{2}+1}^{\uparrow *}(x,\vec{k}'_\perp) \psi_{-\frac{1}{2}+1}^{\uparrow}(x,\vec{k}_\perp) \bigg] \,,
\label{h2}
\eea
\be
\frac{\Delta^1- i \Delta^2}{2 M} E(x,0,-\vec{\Delta}_\perp^2)=\int{\frac{d^2\vec{k}_\perp}{16\pi^3}} \Big[\psi_{+\frac{1}{2}+1}^{\uparrow *}(x,\vec{k}'_\perp) \psi_{+\frac{1}{2}+1}^{\downarrow}(x,\vec{k}_\perp)+\psi_{+\frac{1}{2}-1}^{\uparrow *}(x,\vec{k}'_\perp) \psi_{+\frac{1}{2}-1}^{\downarrow}(x,\vec{k}_\perp) \Big].
\label{e2}
\ee
Using the relation $\vec{k}'_\perp=\vec{k}_\perp-(1-x)\vec{\Delta}_\perp$ and the electron wavefunctions \cite{model1}, we get
\be
E(x,0,-\vec{\Delta}_\perp^2)=- 2 M \left(M-\frac{m}{x}\right)x^2(1-x) I_1 ,
\label{e3}
\ee
\be
I_1= \pi \int_{0}^{1} \frac{d \alpha}{D} , ~~~~~~{\rm and}~~~~~~
D= \alpha (1-\alpha) (1-x)^2 \Delta^2_{\perp}-M^2 x (1-x)+ m^2 (1-x)+\lambda^2 x .
\ee

Since the Fourier transform (FT) diagonalizes the convolution integral, we switch to transverse position space representation of the LCWF by taking Fourier transform in $\vec{\Delta}_\perp$ as
\be
\mathcal{H}(x,\vec{b}_\perp)=\frac{1}{(2 \pi)^2}\int d^2\vec{\Delta}_\perp e^{-i \vec{b}_\perp \cdot \vec{\Delta}_\perp} H(x,0,-\vec{\Delta}_\perp^2),~~~~
\mathcal{E}(x,\vec{b}_\perp)=\frac{1}{(2 \pi)^2}\int d^2\vec{\Delta}_\perp e^{-i \vec{b}_\perp \cdot \vec{\Delta}_\perp} E(x,0,-\vec{\Delta}_\perp^2),
\label{fourierHE}
\ee
where $\vec{b}_\perp$ is the impact parameter conjugate to $\vec{\Delta}_\perp$ and represents the transverse distance between the active quark and the center of mass momentum.
%\begin{figure}
%\minipage{0.42\textwidth}
%    \includegraphics[width=5.8cm,angle=270]{ipdpdf.eps}
%  \endminipage\hfill
%  \minipage{0.42\textwidth}
%  \includegraphics[width=5.8cm,angle=270]{ipdpdf1.eps}
%  \endminipage\hfill
% \caption{Plots of $\mathcal{E}(x,b_\perp)$ as a function of $ b_\perp $ and $x$ for three different values of $x$ and $b_\perp$ respectively.}
%  \label{impactE}
%\end{figure}\\
\begin{figure}
%\minipage{0.42\textwidth}
%   \includegraphics[width=6cm]{a.eps}
%  \endminipage\hfill
  \minipage{0.42\textwidth}
  \includegraphics[width=6cm]{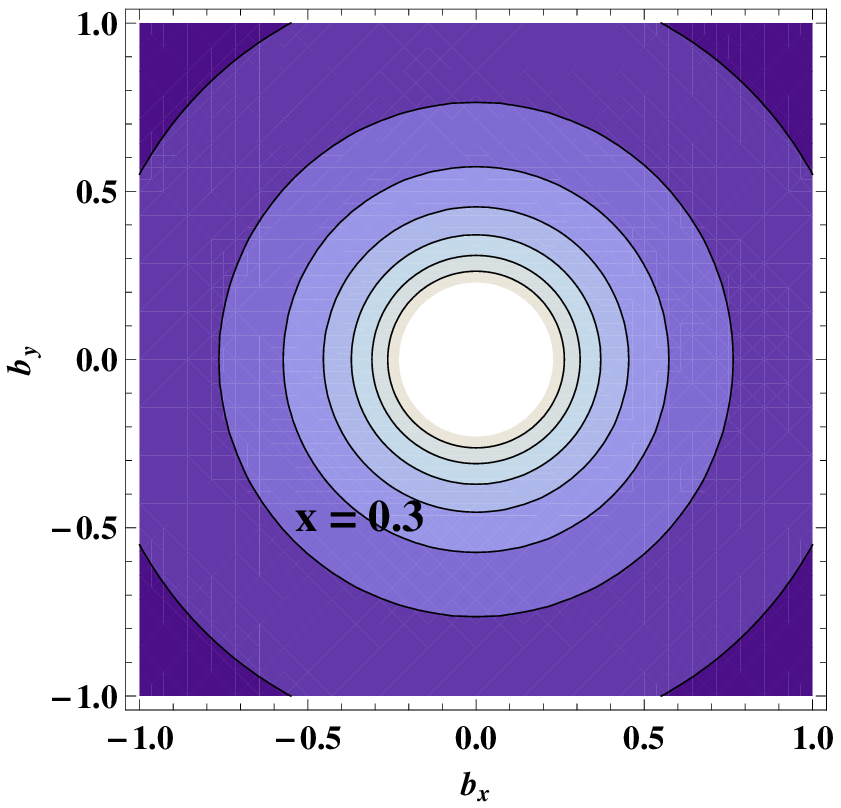}
  \endminipage\hfill
  %\minipage{0.42\textwidth}
%\includegraphics[width=6cm]{c.eps}
% \endminipage\hfill
  \minipage{0.42\textwidth}
  \includegraphics[width=6cm]{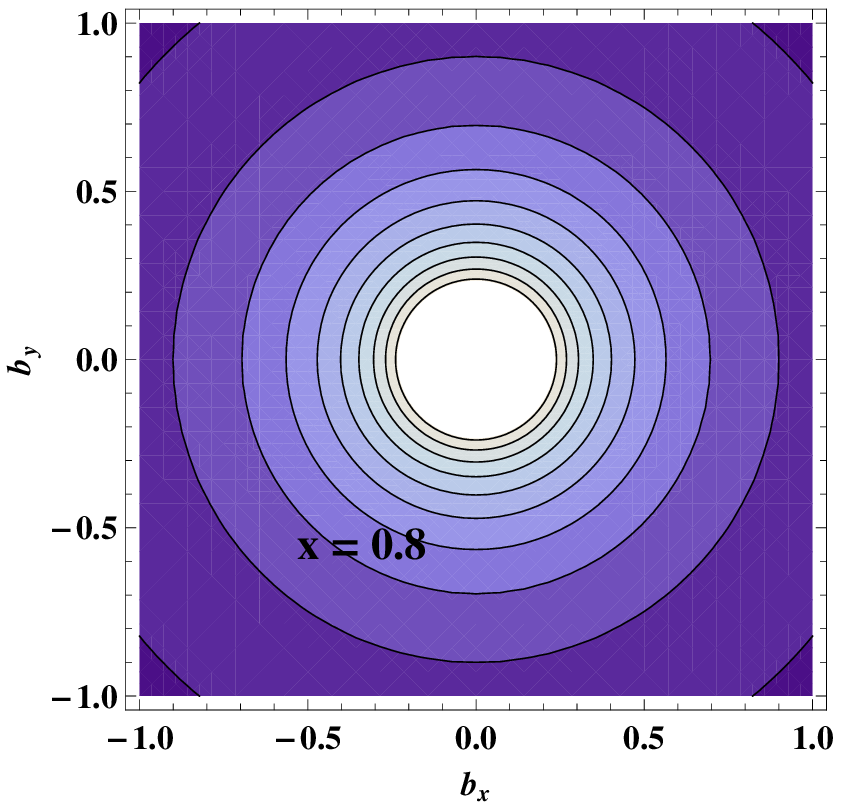}
  \endminipage\hfill
  \caption{Plots of ipdpdf $\mathcal{E}(x,b_\perp)$ for two different values of $x$.}
  \label{contour}
\end{figure}
In Fig. \ref{contour} we present quark distribution in the transverse plane for $x$= 0.3 and 0.8. It describes the quark distribution for the unpolarized nucleon and it is clear from the plot that for lower values of $x$ the distribution is spread over the whole region but as the value of $x$ increases it gets denser near the center.
\section{Transverse distortion of the wave function}

To understand the physical significance of $\mathcal{E}(x,\vec{b}_\perp)$, we consider a state polarized in the $+\hat{y}$ direction with it's transverse center of momentum at the origin $|P^+,\vec{R}_\perp=\vec{0}_\perp,+\hat{y} \rangle=\frac{1}{\sqrt{2}}(|P^+,\vec{R}_\perp=\vec{0}_\perp,\uparrow \rangle +i |P^+,\vec{R}_\perp=\vec{0}_\perp,\downarrow \rangle,$
where $|P^+,\vec{R}_\perp=\vec{0}_\perp,\lambda \rangle= \mathcal{N} \int d^2\vec{P}_\perp |P^+,\vec{P}_\perp,\lambda \rangle.$ $\mathcal{N}$ is the normalization factor and it is chosen such that we get the parton distributions when the impact parameter dependent distributions are integrated over $d^2\vec{b}_\perp$.
The transverse distance from center of momentum can defined using the light cone  momentum density component of the energy momentum tensor and can be expressed as $\vec{R}_\perp\equiv \frac{1}{P^+} \int d^2\vec{x}_\perp \int dx^- T^{++} \vec{x}_\perp = \sum_{i=q,g} x_i \vec{r}_{\perp,i}.$
Here, $x_i$ are the light cone momentum fractions carried by each parton and the sum in the parton representation of $\vec{R}_\perp$ extends over the transverse positions $\vec{r}_{\perp,i}$ of all quarks and gluons in the target. Using the operator $\hat{O}_q(x,\vec{b}_\perp)=\int \frac{dy^-}{4 \pi}\bar{\psi} \left( -\frac{y^-}{2} \vec{b}_\perp \right) \gamma^+ \psi \left( \frac{y^-}{2} \vec{b}_\perp \right) \times e^{ixP^+y^-},$ and the light front gauge $A^+=0$ for a state polarized in $+\hat{y}$ direction, we get the unpolarized quark distribution in impact parameter space \cite{ipd1} expressed as
\be
q_{\hat{y}}(x,\vec{b}_\perp)=\int \frac{d^2 \vec{\Delta}_\perp}{(2\pi)^2} e^{-i \vec{\Delta}_\perp \cdot \vec{b}_\perp} [H(x,0,-\vec{\Delta}_\perp^2)]+i \frac{\Delta^x}{2 M} E(x,0,-\vec{\Delta}_\perp^2)].
\label{e8}
\ee
Using Eq. (\ref{fourierHE}), we get the unpolarized quark distribution in terms of the Fourier transformed GPDs as follows
\be q_{\hat{y}}(x,\vec{b}_\perp)= \mathcal{H}(x,\vec{b}_\perp)+\frac{1}{2 M} \frac{\partial}{\partial b^x} \mathcal{E}(x,\vec{b}_\perp).\ee
It is clear that if the spin flip GPD $E(x,0,t)$ is non zero, the parton distribution of quarks in the transverse plane is distorted for the target having transverse polarization. %A nucleon whose spin is pointing in the positive $\hat{y}$ direction is given by Eq. (11).

%%%%%%%%%%
On the other hand, it is well known that $\mathcal{E}(x,\vec{b}_\perp)$ is a function of $x$ and $\vec{b}_\perp$. When integrated over both $x$ and $\vec{b}_\perp$ it gives the quark contribution to anomalous magnetic moment as $\int dx \int d^2 \vec{b}_\perp \mathcal{E}(x,\vec{b}_\perp)=\kappa.$
The $b^x$ derivative of $\mathcal{E}(x,\vec{b}_\perp)$ is positive for negative $b^x$ and negative for positive $b^x$. Further, the sign of anomalous magnetic moment determines the sign of distortion of quark distribution in impact parameter space. For nucleon which is polarized in the $\hat{y}$ direction, the distortion is towards negative $\hat{x}$ for positive $b^x$ and towards positive $\hat{x}$ for negative $b^x$ when $\kappa>0$ and reverse is the case for $\kappa<0$. We can verify the above assumptions in the present study. The distortion in impact parameter space for a polarized nucleon  is given as
\bea
%\mathcal{E}(x,b_\perp)&=&\int \frac{d^2 \Delta_\perp}{(2 \pi)^2} e^{- \iota b_\perp \cdot \Delta_\perp} E(x,t) \nonumber\\
\frac{\partial}{\partial b^x} \mathcal{E}(x,\vec{b}_\perp)&=& - \frac{1}{2 \pi} \int \Delta^2 J_1(\Delta b) E(x,0,-\vec{\Delta}_\perp^2) d\Delta .
\label{e11}
\eea

In addition to the sign of distortion in impact parameter space, there is infact a more direct way to determine this sign. We take the wavefunctions for a nucleon polarized in $+\hat{y}$ direction and derive
\bea
\psi^{+ \hat{y}}_{+\frac{1}{2}+1}(x,\vec{k}_\perp)=\frac{k^x-i k^y}{x(1-x)} \varphi, &~~~~&
\psi^{+ \hat{y}}_{+\frac{1}{2}-1}(x,\vec{k}_\perp)= - \left(\frac{k^x+ i k^y}{1-x}  + i \left(M-\frac{m}{x}\right)\right)\varphi, \nonumber\\
\psi^{+ \hat{y}}_{-\frac{1}{2}+1}(x,\vec{k}_\perp)= - \left(\left(M-\frac{m}{x}\right) + i \frac{-k^x+ i k^y}{1-x}\right) \varphi, &~~~~&
\psi^{+ \hat{y}}_{-\frac{1}{2}-1}(x,\vec{k}_\perp)= - i \frac{k^x + i k^y}{x (1-x)} \varphi .
\label{e13}
\eea
The unintegrated momentum space distribution, which is even in $k_\perp$, can be obtained by squaring the above equations and we get
\be
q_{\hat{y}}(x,\vec{k}_\perp)= \frac{1}{2 \pi}\left[\frac{k^2_\perp (1+x^2)}{x^2 (1-x)^2}+\left(M-\frac{m}{x}\right)^2\right] \varphi^2 ,
\label{e14}
\ee
where $\varphi$ is supposed to be real.
We can prove explicitly that there is an asymmetry in the $\hat{x}$ direction in the state corresponding to Eq. (\ref{e13}). For this purpose, we perform a Fourier transformation to the transverse position space coordinate say $\vec{f}_\perp$. We now have
\be
\psi^{+ \hat{y}}_{+\frac{1}{2}+1}(x,\vec{f}_\perp) \equiv \int \frac{d^2 k_\perp}{(2 \pi)^2} e^{i \vec{f}_\perp \cdot \vec{k}_\perp} \psi^{+ \hat{y}}_{+\frac{1}{2}+1}(x,\vec{k}_\perp) = \frac{1}{x(1-x)}\left(- i \frac{\partial}{\partial f^x}-\frac{\partial}{\partial f^y}\right) \varphi(f_\perp) ,
\ee
\be
\psi^{+ \hat{y}}_{+\frac{1}{2}-1}(x,\vec{f}_\perp)\equiv \int \frac{d^2 k_\perp}{(2 \pi)^2} e^{i \vec{f}_\perp \cdot \vec{k}_\perp} \psi^{+ \hat{y}}_{+\frac{1}{2}-1}(x,\vec{k}_\perp) = -\left[\frac{1}{(1-x)}\left(- i \frac{\partial}{\partial f^x}+\frac{\partial}{\partial f^y}\right)+ i \left(M-\frac{m}{x}\right)\right]\varphi(f_\perp) ,
\ee
\be
\psi^{+ \hat{y}}_{-\frac{1}{2}+1}(x,\vec{f}_\perp)\equiv \int \frac{d^2 k_\perp}{(2 \pi)^2} e^{i \vec{f}_\perp \cdot \vec{k}_\perp} \psi^{+ \hat{y}}_{-\frac{1}{2}+1}(x,\vec{k}_\perp) = -\left[\left(M-\frac{m}{x}\right)+\frac{i}{1-x}\left(i \frac{\partial}{\partial f^x}+\frac{\partial}{\partial f^y}\right)\right]\varphi(f_\perp) ,
\ee
\be
\psi^{+ \hat{y}}_{-\frac{1}{2}-1}(x,\vec{f}_\perp)\equiv \int \frac{d^2 k_\perp}{(2 \pi)^2} e^{i \vec{f}_\perp \cdot \vec{k}_\perp} \psi^{+ \hat{y}}_{-\frac{1}{2}-1}(x,\vec{k}_\perp)
= \frac{- i}{x(1-x)} \left(-i \frac{\partial}{\partial f^x}+\frac{\partial}{\partial f^y}\right)\varphi(f_\perp) ,
\ee
where
\be
\varphi(\vec{f}_\perp)\equiv  \int \frac{d^2 k_\perp}{(2 \pi)^2} e^{i \vec{f}_\perp \cdot \vec{k}_\perp} \varphi(\vec{k}_\perp) = - e x \sqrt{1-x}  \int \frac{d^2 k_\perp}{(2 \pi)^2} e^{i \vec{f}_\perp \cdot \vec{k}_\perp} \frac{1}{k^2_\perp + C} = -\frac{e}{2 \pi} x \sqrt{1-x} K_0(|f_\perp \sqrt{C}|),
\ee
and
\be
C=m^2(1-x)- M^2 x (1-x)+ \lambda^2 x.
\ee
\begin{figure}
%\minipage{0.42\textwidth}
%   \includegraphics[width=6cm]{q1.eps}
%  \endminipage\hfill
  \minipage{0.42\textwidth}
  \includegraphics[width=6cm]{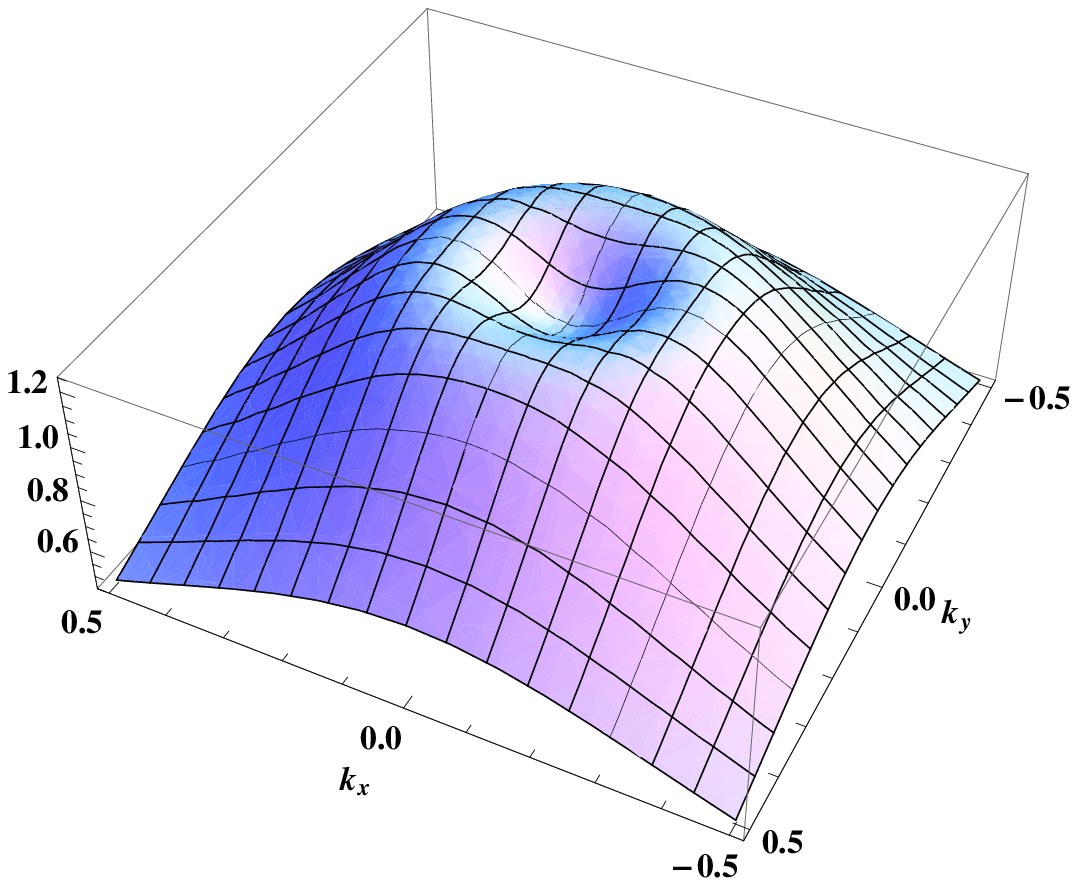}
  \endminipage\hfill
 % \minipage{0.42\textwidth}
%\includegraphics[width=5cm]{q9.eps}
%  \endminipage\hfill
  \minipage{0.42\textwidth}
  \includegraphics[width=6cm]{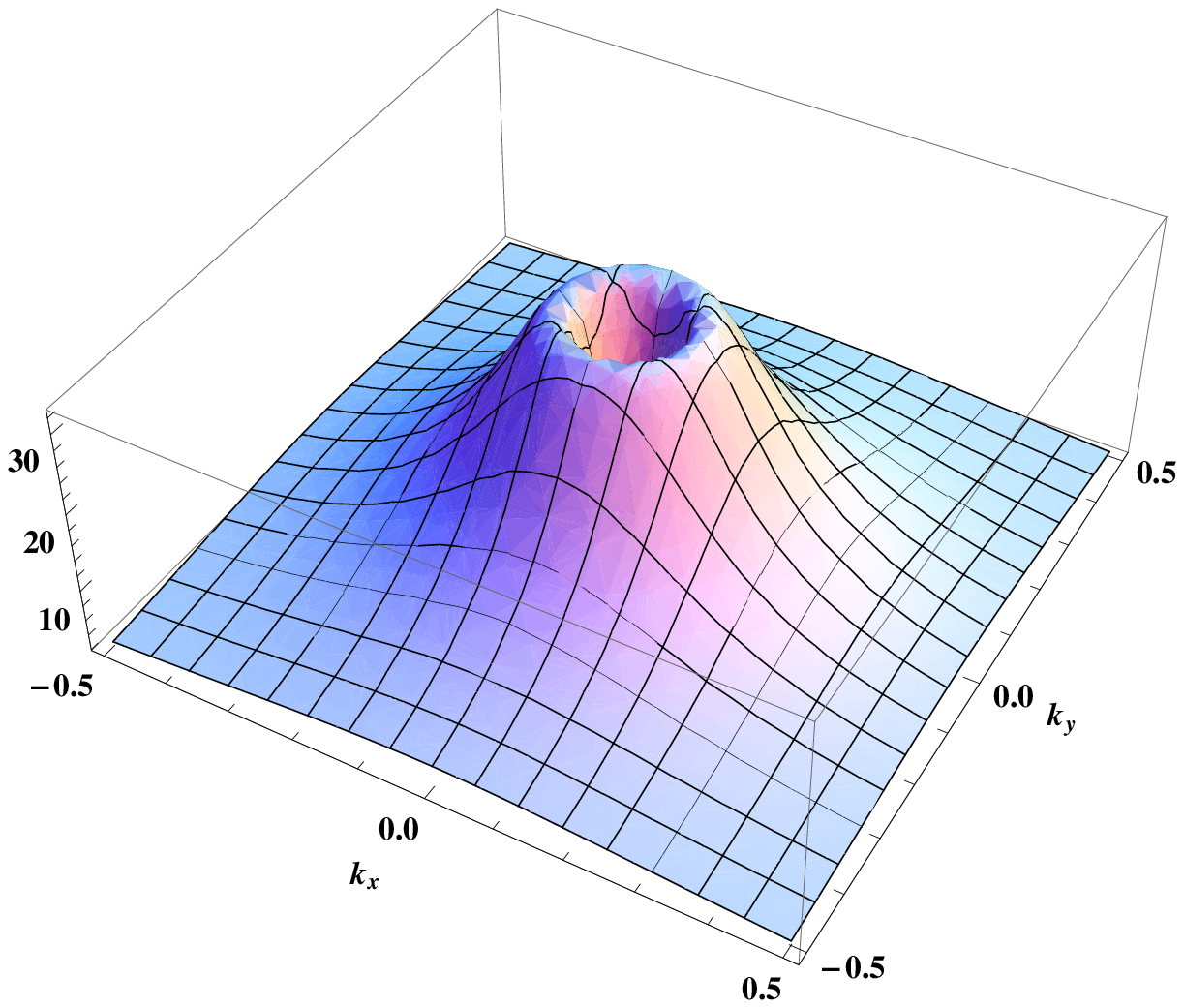}
  \endminipage\hfill
  \caption{Plot of $q_{\hat{y}}(x,\vec{k}_\perp)$ vs $k_\perp$ for two different values of $x=(0.4,0.8)$ .}
  \label{3d-distribution}
\end{figure}
%Using the relations
%\begin{eqnarray*}
%|\psi^{+ \hat{y}}_{+\frac{1}{2}+1}(x,\vec{f}_\perp)|^2 &=& |\psi^{+ \hat{y}}_{-\frac{1}{2}-1}(x,\vec{f}_\perp)|^2, \nonumber\\
%|\psi^{+ \hat{y}}_{+\frac{1}{2}-1}(x,\vec{f}_\perp)|^2 &=& |\psi^{+ \hat{y}}_{-\frac{1}{2}+1}(x,\vec{f}_\perp)|^2,
%\label{e15}
%\end{eqnarray*}
The unpolarized quark distribution in transverse coordinate space $\vec{f}_\perp$ can be expressed as
\bea
q_{\hat{y}}(x,\vec{f}_\perp)
&=&\frac{1}{2}\left[\frac{(1+x^2)}{x^2 (1-x)^2}\left(\frac{\partial}{\partial f^x} \varphi \right)^2+ \frac{1+x^2}{x^2 (1-x)^2}\left(\frac{\partial}{\partial f^y} \varphi\right)^2+ \left(M-\frac{m}{x}\right)^2 \varphi^2\right]\nonumber\\
&&-\left(M-\frac{m}{x}\right)\varphi \frac{1}{1-x}\left(\frac{\partial}{\partial f^x} \varphi \right).
\label{e16}
\eea
In Fig. \ref{3d-distribution} we present the unintegrated momentum space distribution obtained from the light front wavefunctions for different values of $x$, to check the sign of distortion for a polarized nucleon in impact parameter space. It is clear from the plots that at lower value of $x$ there is maxima at origin but as the value of $x$ is increased some distortion is observed. As the value at $x$ is further increased the distortion also increases towards negative direction. These plots helps us to determine the distortion sign directly from the light front wavefunctions.

To summarize, in the present work we have studied the GPDs in impact parameter space obtained from LFWFs. We have shown that if spin flip GPD is non-zero then parton distribution is distorted in the transverse plane when the target nucleon has transverse polarization. The  sign of the distortion is obtained from the sign of anomalous magnetic moment. We have considered the nucleon polarized in $+\hat{y}$ direction and obtained the unintegrated momentum space distribution directly from the LCWFs and is found to be even in $k_\perp$.

\end{document}